\documentclass[conference]{IEEEtran}
\usepackage{cite}
\usepackage{amsmath,amssymb,amsfonts}
\usepackage{algorithmic}
\usepackage{graphicx}
\usepackage{textcomp}
\usepackage{xcolor}
\usepackage{booktabs}
\usepackage{caption}
\usepackage{subcaption}
\usepackage{array}
\usepackage[a4paper, total={184mm,239mm}]{geometry}
\renewcommand{\baselinestretch}{0.99}
\def\BibTeX{{\rm B\kern-.05em{\sc i\kern-.025em b}\kern-.08em
    T\kern-.1667em\lower.7ex\hbox{E}\kern-.125emX}}
\begin{document}
\newcommand{\fuseconv}{FuSeConv}
\title{FuSeConv: Fully Separable Convolutions for Fast Inference on Systolic Arrays
}
\author{\IEEEauthorblockN{Surya Selvam, Vinod Ganesan, Pratyush Kumar}
\IEEEauthorblockA{Department of Computer Science and Engineering, IIT Madras, India}
\IEEEauthorblockA{selvams@purdue.edu, \{vinodg,pratyush\}@cse.iitm.ac.in}}
\maketitle
\begin{abstract}
Both efficient neural networks and hardware accelerators are being explored to speed up DNN inference on edge devices.
For example, MobileNet uses depthwise separable convolution to achieve much lower latency, while systolic arrays provide much higher performance per watt.
Interestingly however, the combination of these two ideas is inefficient: The computational patterns of depth-wise separable convolution are not systolic and lack data reuse to saturate the systolic array's constrained dataflow. 
In this paper, we propose FuSeConv (Fully-Separable Convolution) as a drop-in replacement for depth-wise separable convolution.
FuSeConv generalizes the decomposition of convolutions fully to separable 1D convolutions along spatial and depth dimensions. 
The resultant computation is systolic and efficiently utilizes the systolic array with a slightly modified dataflow.
With FuSeConv, we achieve a significant speed-up of 3x-7x with the MobileNet family of networks on a systolic array of size 64x64, with comparable accuracy on the ImageNet dataset. 
The high speed-up motivates exploration of hardware-aware Neural Operator Search (NOS) in complement to ongoing efforts on Neural Architecture Search (NAS).

\end{abstract}


\section{Introduction}
Deep Neural Networks (DNNs) continue to establish state-of-the-art accuracy on tasks across domains.
However, as accuracy requirements create larger and more complex DNNs, efficient inference on these networks becomes the primary challenge. 
There are two broad approaches to address this challenge - domain-specific hardware \textit{accelerators} and efficient \textit{operators} in DNNs.
In many hardware accelerators, a systolic array \cite{kung1982systolic} is a popular design pattern to accelerate matrix multiplication and convolution with a grid of multiply-accumulate units (MACs).
It's efficiency is exemplified by TPUs \cite{jouppi2017datacenter}: At the time of release, the TPUv1 provided 25 to 29 times higher performance-per-watt than comparable GPUs \cite{jouppi2017datacenter}.
On the other hand, depthwise separable convolution is a good example of an optimized operator in DNNs.
A depthwise separable convolution decomposes a standard convolution into independent convolutions on each input channel, called depthwise convolution, followed by a 1x1 pointwise convolution.
This reduces the number of parameters and operations as seen in the MobileNet \cite{howard2017mobilenets, sandler2018mobilenetv2, howard2019searching} family of networks: MobileNet-V3 has 14.5 times fewer MACs but 1.6\% higher accuracy on ImageNet than DenseNet-121 \cite{huang2017densely}.

Surprisingly however, the combination of these two successful ideas, {\em i.e.}, systolic arrays executing depthwise separable convolution, does not work as expected. 
For instance, MobileNet-V2 has 12$\times$ fewer computations than ResNet-50, but runs only 1.3$\times$ faster on a systolic array with MACs arranged in a $32 \times 32$ array. 
This incommensurate scaling has been identified earlier on EdgeTPU running EfficientNet \cite{gupta2020accelerator} and while designing SqueezeNext \cite{gholami2018squeezenext}.
In this work, we look at this discrepancy more closely by posing three research questions: 
Formally, why does depthwise separable convolution not work as well on systolic arrays?
Can we design a drop-in replacement for it that is fast on systolic arrays?
How good is such a replacement in terms of execution time and accuracy?

\begin{figure*}
    \centering
    \vspace*{-20pt}
    \includegraphics[width=0.9\textwidth]{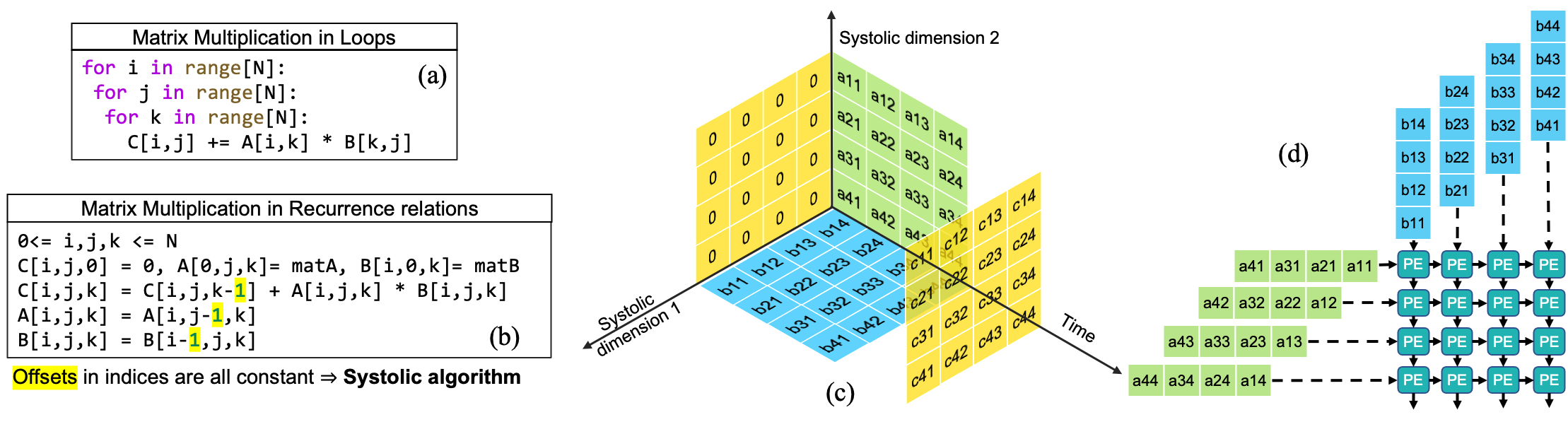}
    \caption{(a). Loop representation of matrix multiplication; (b). Corresponding recurrence relations; (c). Geometric representation of the recurrence relation; (d). Output-stationary dataflow mapping to a systolic-array}
    \label{fig:mm}
    \vspace*{-15pt}
\end{figure*}

For the first question, we study the formalism of Regular Iterative Algorithms (RIA) \cite{rao1988regular} and show that 2D convolution is not a {\em systolic algorithm} -- a class of algorithms that efficiently run on systolic architectures \cite{song1994systolic}.
Consequently, to map depth-wise seprabable convolution on to a systolic array, we need to apply the im2col \cite{chellapilla2006high} transformation. 
We show that this transformed computation does not have any data reuse on 2D systolic arrays, and thus has very poor utilization.

For the second question, we propose Fully Separable convolution (FuSeConv) as a drop-in replacement for depthwise separable convolution. 
FuSeConv generalizes the decomposition fully to separable 1D convolutions along all spatial and depth dimensions.
1D convolutions are systolic algorithms and thus can be efficiently mapped on to systolic arrays.
We show that for efficient execution on 2D systolic arrays, a small change in the dataflow is required - each row should support a weight-broadcast link.
We evaluate that the cost of this additional dataflow pattern is small: For instance, for a systolic array of size $32\times 32$, only $4.35\%$ area and $2.25\%$ power overheads are incurred when synthesized on a 45nm node.

For the third question, we extensively evaluate networks with FuSeConv layers for the ImageNet dataset. 
In particular, we replace all depthwise convolution layers in MobileNet(V1,2,3) and MNasNet networks with two proposed variants of FuSeConv layers.
With detailed evaluation we show that one of the variants is more effective and achieves significant speed-ups on all three versions of MobileNet and MNasNet.
In particular, we achieve a speed-up of \textbf{$6.76\times$} on MobileNetV1 on a systolic array of size $64\times64$, while matching accuracy on ImageNet.
Similarly on MobileNetV2 we obtain a speed-up of $5.1\times$.

In summary, our work proposes a different primitive operator, fully decomposed 1D convolutions, for executing DNNs on systolic arrays.
The significant out-performance of this proposed operator relative to already efficient networks motivates greater focus on hardware-aware Neural \textit{Operator} Search (NOS). 
We see NOS as a natural complement to ongoing active research on Neural Architecture Search (NAS), since it informs the choice of operators considered in the search space of NAS.

The rest of the paper is organized as follows. 
We discuss background material in Section~\ref{sec:background}.
In Section~\ref{sec:depthwise}, we show why depthwise convolution is not efficient on systolic arrays. 
We propose FuSeConv and a modified dataflow in Section~\ref{sec:proposed}. 
We present experimental results in Section~\ref{sec:experimental} and conclude in Section~\ref{sec:conclusion}.

\label{sec:introduction}

\section{Background}
\label{sec:background}
\noindent
\subsection{Systolic Arrays}

A systolic array \cite{kung1982systolic} defines a regular arrangement, such as a rectangular grid, of homogeneous processing elements (PEs).
The word `systolic' implies rhythmic patterns in communication and computation, which are globally synchronous.
This constrained dataflow restricts applications that can be mapped onto systolic arrays.
However, the ones that can be mapped, such as matrix multiplication, have improved performance due to low control overhead and main memory dependence.
Systolic arrays are used in many DNN accelerators \cite{jouppi2017datacenter, XlinixDNNProcessor:online, bannon2020accelerated}.


\subsection{Systolic Algorithms} 

We discuss systolic algorithms with the example of matrix multiplication, as studied in the pioneering work of Kung and Leiserson \cite{kung1980systolic}.
In Fig.~\ref{fig:mm}(a) we show the standard loop implementation of a matrix-matrix multiplication. 
We transform this into the format of recurrence relations as shown in Fig.~\ref{fig:mm}(b).
These relations are said to define a \textit{regular iterative algorithm} (RIA) \cite{rao1988regular}, since they satisfy three conditions: 
(a) Each variable is defined by a name and a set of indices (in this case 3). 
(b) Each variable is assigned a value just once (single assignment language). 
(c) For each recurrence relation, the difference between the indices of the variable in the LHS and each variable in the RHS is a constant.
For instance, in the equation for $C[i, j, k]$, the differences in indices (called index offsets) to the three variables $A$, $B$, and $C$ in the RHS are $[0, 0, 0], [0, 0, 0], $ and $[0, 0, -1]$, respectively.
RIAs that satisfy these three conditions are a super-set of algorithms which can be synthesized on systolic arrays, \textit{i.e.}, systolic algorithms \cite{wan1996systolic}.

\subsection{Mapping systolic algorithms on to systolic arrays}
The computation of the said recurrence relations can be visualized geometrically as shown in Fig.~\ref{fig:mm}(c).
Each point in the 3D space corresponds to a combination of the indices $i, j, k$ starting at the origin $(0,0,0)$.
As shown, inputs $A$ and $B$ are initialized in respective planes while output $C$ is initialized with 0s. 
All values are propagated through to other points as per the recurrence relations. 
The computations for updating the values are mapped on to three dimensions.
The dimensions $i$ and $j$, along which $A$ and $B$ are propagated, are marked as systolic dimensions. 
The dimension $k$ along which $C$ is updated is marked as the `time' dimension, ending with the final computed value as shown. 
Assigning such dimensions is equivalent to \textit{mapping} the algorithm on to a 2D systolic array as shown in Fig.~\ref{fig:mm}(d).
In the mapping, matrices $A$ and $B$ are input along rows and columns (the two systolic dimensions), respectively. 
The output $C$ is computed in each processing element (PE) over time (the time dimension).
Since the output remains stationary in the PEs, this dataflow is referred to as \textit{output stationary}. 
We can similarly study input and weight stationary dataflows.

\subsection{Convolution Operations}
In \textit{standard convolution}, an input of size $W\times H\times C$ is convolved with a filter of size $K\times K\times C$ to obtain an output of size $N\times M\times 1$, where $N=W-K+1$ and $M=H-K+1$. 
The output of $C'$ convolution filters are stacked to obtain an output of size $N\times M\times C'$.
In \textit{depthwise convolution}, an input of size $W\times H\times C$ is convolved with a filter of size $K\times K\times C$ channel-wise, \textit{i.e.}, every $W\times H$ channel is convolved with the respective $K\times K$ channel in the filter to obtain an output of size $N\times M\times C$.
This is followed by a \textit{point-wise convolution} with $C'$ filters of size $1\times 1\times C$ to obtain an output of size $N\times M\times C'$.
For both these illustrated cases, the input and output sizes are the same.
However, there is a major difference in the number of operations: Standard convolution has a total of $NMC'K^2C$ operations, while depthwise separable convolution has $NMC(K^2+C')$ operations.
This reduction in number of operations with depthwise separable convolution translates to much reduced inference time at comparable accuracy values.

\section{Why Depthwise convolutions are inefficient on Systolic arrays?}
\label{sec:depthwise}
In this section, we show why depthwise separable convolution has poor scaling performance on systolic arrays.
We also explain this finding in contrast to the widespread use of systolic arrays for standard convolution.


\subsection{2D convolutions are not systolic algorithms}

\begin{figure}
    \vspace*{-20pt}
    \centering
    \includegraphics[width=0.45\textwidth]{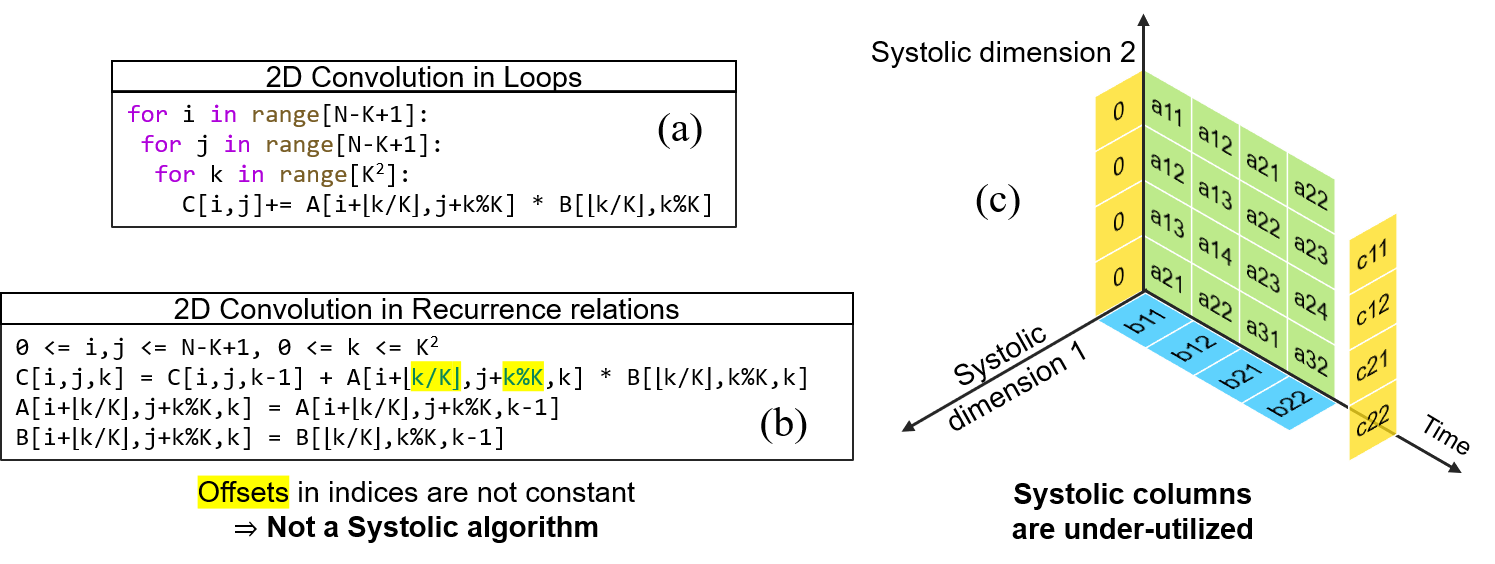}
    \caption{(a) Loop representation of 2D convolution; (b) Corresponding recurrence relations; (c) Geometric representation after transforming 2D convolution to a systolic algorithm.}
    \vspace*{-18pt}
    \label{fig:convsysalgo}
\end{figure}

As described earlier, depthwise convolution requires independent convolutions between 2D slices of the input with 2D kernels, henceforth referred to as a 2D convolution. 
2D convolution can be written in loops as shown in Fig.~\ref{fig:convsysalgo}(a): $A$ is the input feature map, $B$ the weight kernel, and $C$ the output.
Is this a systolic algorithm?
We first attempt to transform it into a Regular Iterative Algorithm (RIA) which as a set contains all systolic algorithms.
Fig.~\ref{fig:convsysalgo}(b) shows the recurrence relations for 2D convolution. 
Like in the case of matrix multiplication, we have added a third index to satisfy the single assignment property.
Unlike the case of matrix multiplication, however, we observe that the index offsets between LHS and RHS are not constants.
For instance, in the recurrence relation for $C$, the index offset to $A$ is given as $[\lfloor k/K \rfloor, k\%K, 0]$.
Since this index offset depends on the index $k$, it violates the important requirement for a RIA.

Can this specification be refactored in some other way to satisfy RIA's requirement?
Note that computing the output at index $(i, j)$ requires summing up $K^2$ products.
We can map these computations to $K^2$ values of the $k$ index.
Computing a sum of these products implies a single offset dependence between $C[i, j, k]$ and $C[i, j, k+1]$.
However, for matrices $A$ and $B$, the computation of these products requires input across a grid of $K \times K$ values.
Independent of the order in which these values are accessed, their $i, j$ indices will depend on the index $k$.
However, all these products are summed to the same $i, j$ index of $C$.
Thus, in the same recurrence relation, the $i, j$ index of $C$ remain constant while those of $A, B$ depend on $k$, violating the criterion for constant index offsets.
We thereby conclude that 2D convolution cannot be written as an RIA, and consequently depthwise convolution is not a systolic algorithm.

Though depthwise convolution is not a systolic algorithm, standard convolution operations are mapped on to systolic arrays.
We discuss two ways in which such mapping is done and why the results do not extend to depthwise convolution.

\subsection{Transformation im2col and Data Reuse}
\begin{figure}
    \vspace*{-20pt}
    \centering
    \includegraphics[width=0.45\textwidth]{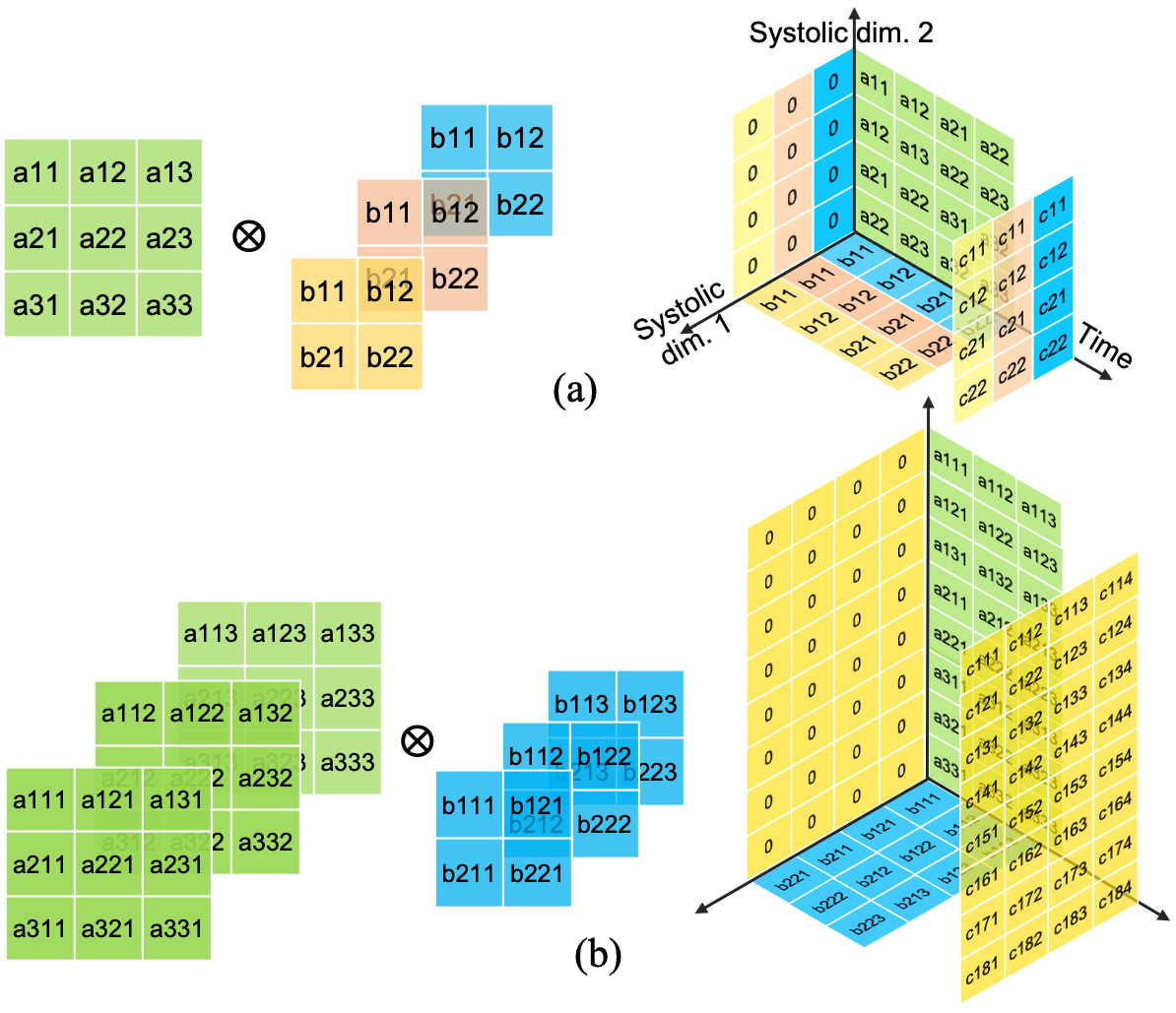}
    \caption{Two methods that enable execution of standard convolution on systolic arrays: (a) Input reuse across filters, (b) Ordering operations along channels.}
    \label{fig:channelwise}
    \vspace*{-15pt}
\end{figure}

Consider a transformation of $A$ such that each set of $K\times K$ values required in each step of convolution is stored in a row.
With this transformation, the index offsets between $C$ and $A$ become constant.
This is the approach with im2col \cite{jia2014caffe} which creates a larger matrix $A'$ from $A$ with repeating entries and a flattened $B$ matrix.
The 2D convolution operation on these modified matrices is a systolic algorithm as shown in Fig.~\ref{fig:convsysalgo}(c).
Note that this computation does not scale on systolic dimension 1, {i.e.}, when mapped to a 2D systolic array it would only use a single column resulting in very poor utilization.
This also implies that data written on to the systolic arrays are not reused across operations.
This lack of data reuse significantly lowers performance of depthwise convolution on systolic arrays. 
However, this is the not the case with standard convolution, where the same input channel has to be convolved with weight matrices from multiple filters enabling data reuse.
This is shown in Fig.~\ref{fig:channelwise}(a), wherein the filters scale along systolic dimension 1 achieving high utilization.
Thus, depthwise convolution's channel-wise decomposition which was designed for higher efficiency also lowers utilization on systolic arrays.



\subsection{Channel-wise operations}
To avoid the expensive im2col transformation, an alternative approach maps standard convolution into channel-wise operations on systolic arrays \cite{jouppi2017datacenter}.
Specifically, standard convolution is implemented as a dot product of vectors of size $C$ along channels from the input and filter matrices.
The corresponding mapping on to a systolic array is shown in Fig.~\ref{fig:channelwise}(b).
The generated output needs to be reduced with an adder tree (usually a part of most systolic array accelerators) to obtain the final output.
In this case too, we are able to utilize both systolic dimensions.
Again, depthwise convolution does not expose any computation spanning channels to benefit from this mapping. 

In summary, with the RIA formalism, we showed that 2D convolution is not a systolic algorithm. 
Further, the two methods of im2col transformation and channel-wise operations, which enable standard convolution to execute on systolic arrays, are not applicable to depthwise convolution.
This explains the poor utilization of systolic arrays with depthwise convolution, and motivates a new systolic-friendly operator.

\section{FuSeConv: Our HW/SW Co-design solution}
\label{sec:proposed}

In this section, we discuss the proposed \fuseconv-operation and how it can be mapped on to a systolic array.



\subsection{The \fuseconv~Operator}

Though depthwise separable convolution decomposes standard convolution to reduce operations, the resultant algorithm is not systolic and does not benefit from either data reuse or channel-wise mapping.
Our motivation with the \fuseconv~is to combine the benefits of decomposing convolution with that of a systolic algorithm.
In addition, we also draw inspiration from the work on grouped convolution \cite{krizhevsky2012imagenet}.

\begin{figure}[t]
    \vspace*{-20pt}
    \centering
    \includegraphics[width=0.44\textwidth]{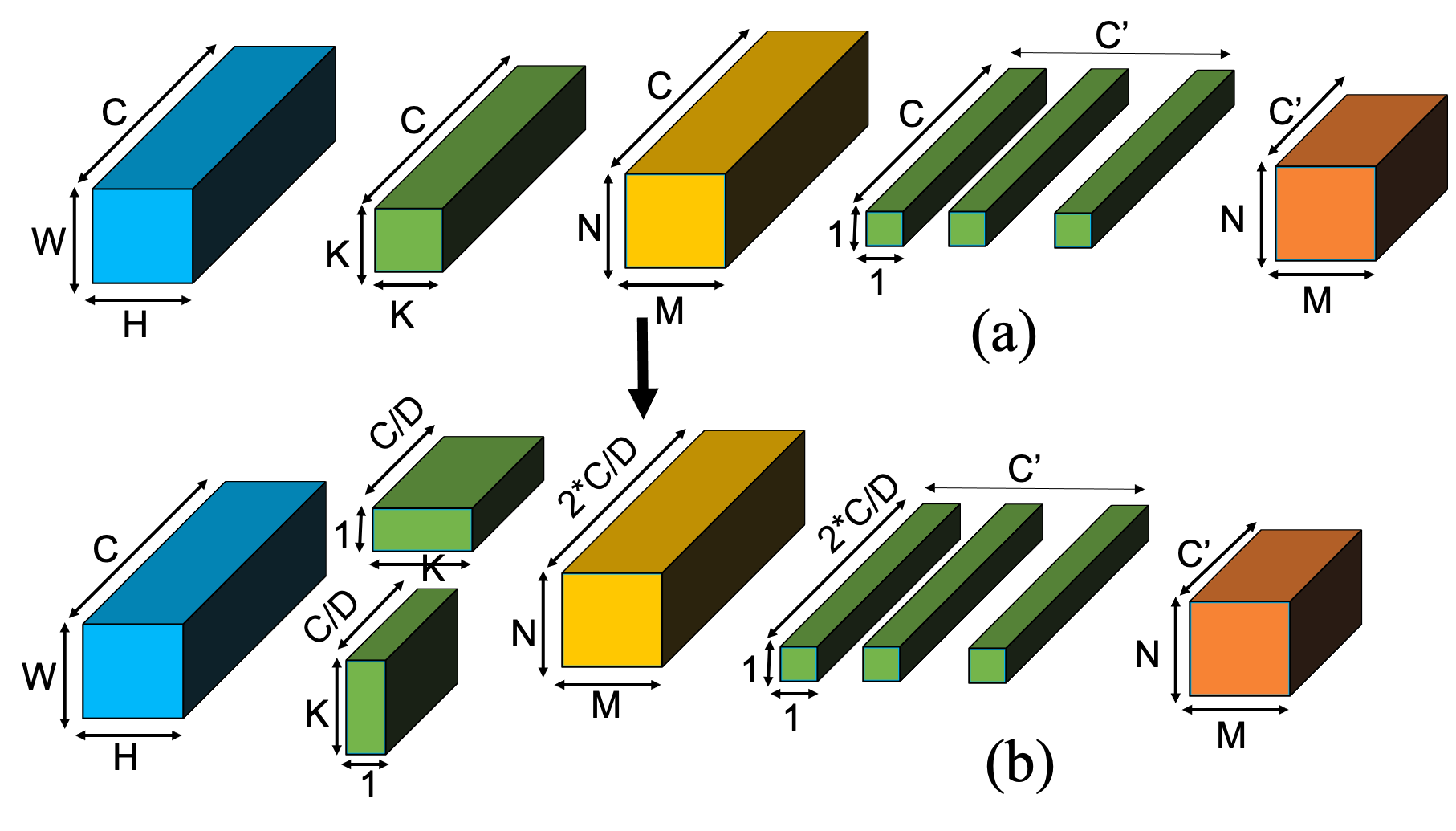}
    \caption{Transformation of a depthwise separable convolution layer into a \fuseconv~layer}
    \label{fig:fuselayer}
    \vspace*{-16pt}
\end{figure}

Recall that a depthwise separable convolution has $K\times K$ 2D convolutions for filtering $C$ input channels independently, followed by $C'$ pointwise filters (see Fig.~\ref{fig:fuselayer}(a)). 
We extend this decomposition with \fuseconv, where we factorize the $K\times K\times C$ depthwise filters into two groups of depthwise filters: $K\times1\times C/D$ 1D row filters and $1\times K\times C/D$ 1D column filters. 
Here, $D$ is a design-knob used to generate variants of \fuseconv~layers. 
The resultant output is passed through $C'$ pointwise filters (see Fig.~\ref{fig:fuselayer}(b)).
In essence, \fuseconv~performs 1D convolutions alone to fully separate the filtering of information along the three axes of the input, and hence the name.



With the same input and output sizes as with depthwise separable convolution, \fuseconv~is designed as a drop-in replacement.
With this replacement, number of parameters changes from $C(K^2+C')$ to $\frac{2}{D} C(K+C')$ and number of operations from $NMC(K^2 + C')$ to $\frac{2}{D} NMC(K + C')$.
To study the trade-off between efficiency and accuracy, we consider two variants for $D = 1$ and $2$. 
In the \textit{full variant} with $D=1$, we apply both row and column filters on all channels for an output of size $K\times K\times 2C$. 
In the \textit{half variant} with $D=2$, row filters operate on $C/2$ channels while column filters operate on the other $C/2$ channels for an output of size $K\times K\times C$.
Clearly, the full variant with $D = 1$ has more parameters and operations.

\subsection{\fuseconv~is a systolic algorithm}
\fuseconv~comprises of independent 1D convolutions which have been extensively studied and found to be systolic algorithms \cite{quinton1984automatic}.
Indeed \cite{kung1982systolic} illustrates 7 different ways of mapping a 1D convolution on to a linear systolic array.
In Fig.~\ref{fig:1Dconv} (a), we show the 1D convolution both as a loop and recurrence relations.
The recurrence relations satisfy the requirements of an RIA. 
The other operation in a \fuseconv~layer, point-wise convolution, is a vector dot-product and is also a systolic algorithm.
Thus, \fuseconv~can be efficiently mapped to systolic arrays without requiring transformations such as im2col.

\subsection{Proposed Hardware Architecture and Mapping}
\begin{figure}[h!]
    \vspace*{-15pt}
    \centering
    \includegraphics[width=0.33\textwidth]{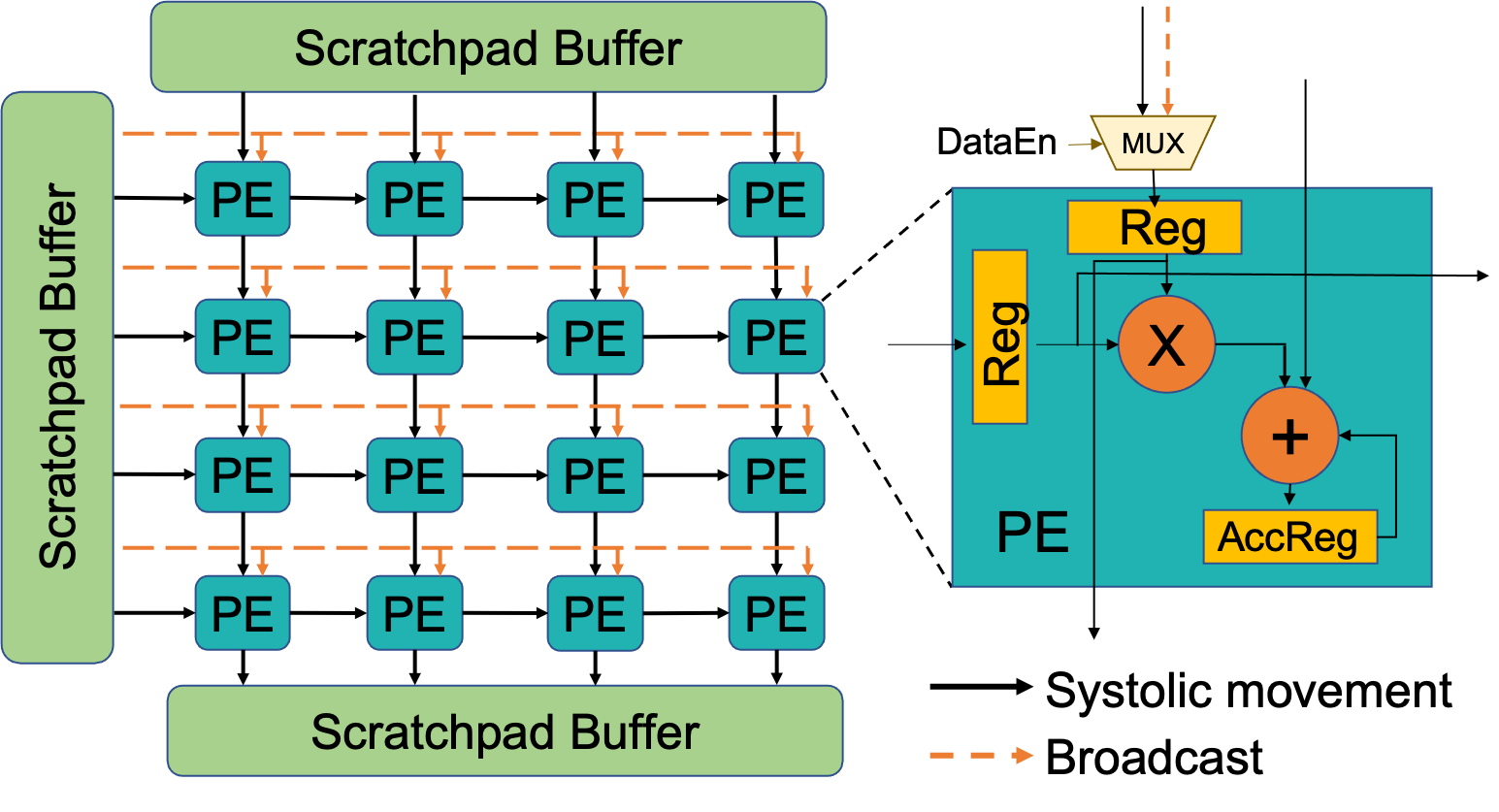}
    \caption{Overview of systolic array with the proposed dataflow}
    \label{fig:systolic}
\end{figure}

\subsubsection{Optimized dataflow for \fuseconv}
The independent 1D convolutions of \fuseconv~can be mapped into individual rows of a 2D systolic array. 
However, this mapping requires a slightly modified dataflow: 
Each row of a systolic array should have a broadcast link that sends weight values to all PEs in that row, which is similar to Eyeriss \cite{chen2016eyeriss} that relies on an NoC instead of a systolic architecture.
This modified dataflow can co-exist with the standard systolic flow from top to bottom. 
For this, the PE can be configured (as shown in Fig.~\ref{fig:systolic}) to either read data from the top systolic link or the row broadcast link.
We compute and report the additional overhead of this modified dataflow in the experimental section.

\begin{figure}
    \vspace*{-15pt}
    \center
    \includegraphics[width=0.5\textwidth]{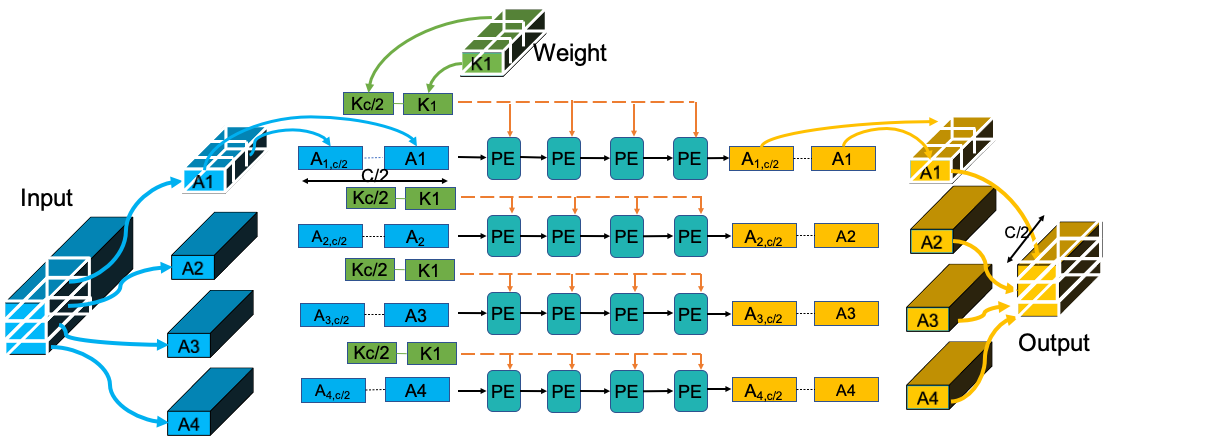}
    \caption{Mapping FuSe layers to the proposed systolic-array architecture}
    \label{fig:mapping}
    \vspace*{-18pt}
\end{figure}


\subsubsection{Mapping \fuseconv~layers}
Fig.~\ref{fig:mapping} illustrates how a \fuseconv~is mapped onto a systolic array of size $S\times S$ with the modified dataflow.
We choose the half variant (\textit{i.e.}, $D=2$), and only show the mapping of row 1D filters. 
The mapping of column filters and the full variant follows similarly and the 1x1 pointwise convolution of \fuseconv~is mapped to the standard systolic dataflow.
The input is sliced into its $W$ rows, denoted $A_1$ through $A_W$, and each slice is allocated to one row of the systolic array.
Every row is further sliced across the channels into $C/2$ channel slices denoted $A_{1,1}$ through $A_{1,C/2}$.
Similarly, weights are sliced across channels into $C/2$ 1D filters denoted $K_1$ through $K_{C/2}$.
The computation follows multiple folds wherein at every fold one weight channel slice operates over one input-channel slice generating $S$ output feature map slices.
After all folds are computed, the output slices are concatenated to form the output feature map.

\begin{figure}
    \vspace*{-15pt}
    \centering
    \includegraphics[width=0.5\textwidth]{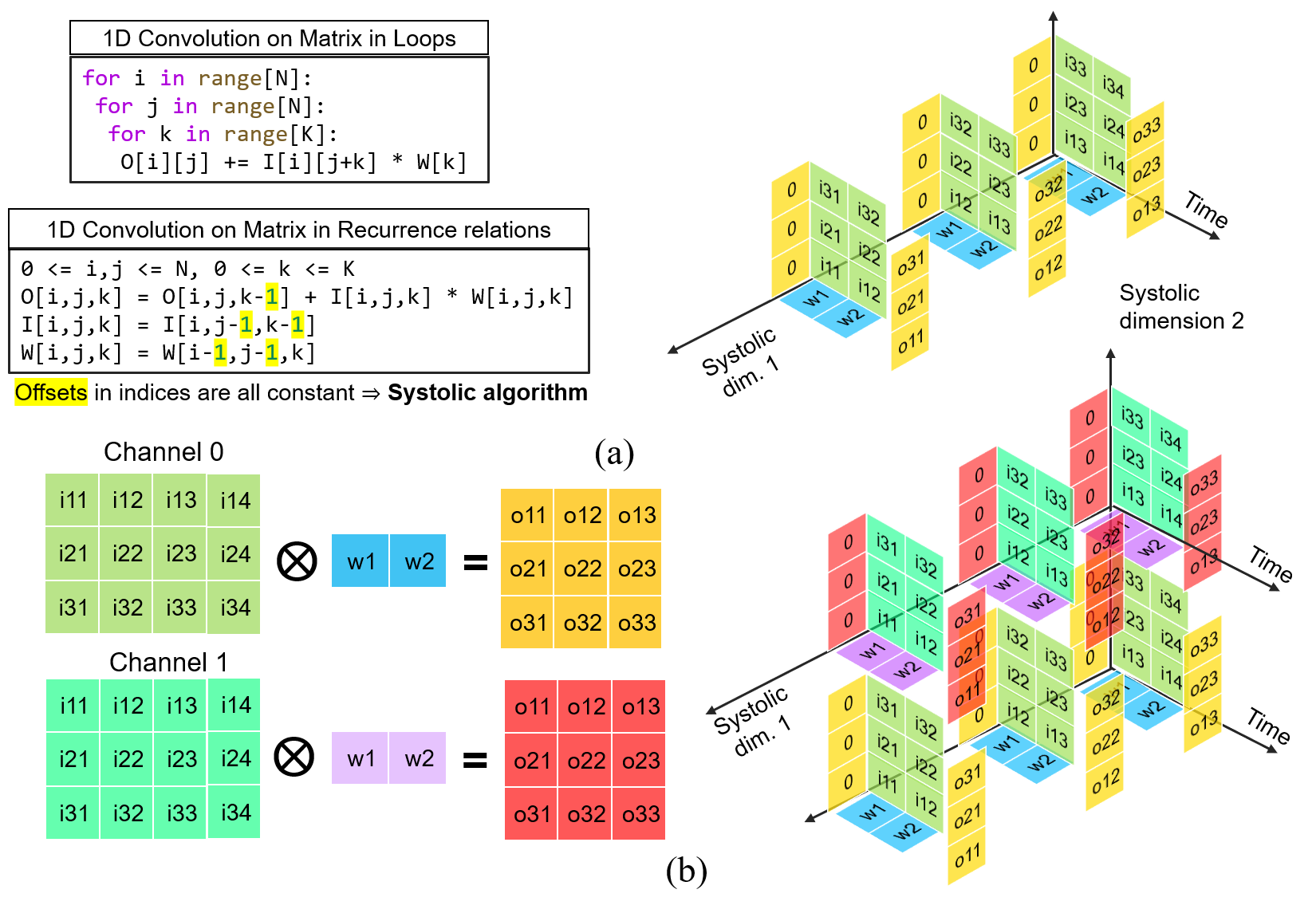}
    \caption{(a) Loop representation, recurrence relation and geometric representation of a 1D convolution; (b) Mapping multiple channels of 1D convolutions to the proposed systolic-array}
    \label{fig:1Dconv}
    \vspace*{-20pt}
\end{figure}

\subsubsection{Efficient utilization}
With the proposed operation, modified dataflow, and mapping, is a systolic array efficiently utilized?
In Fig.~\ref{fig:1Dconv} (a), we visualize a 1D filter of 2 weights convolved with a 4x3 input. 
Note that due to the broadcast link, at any time step the weight value is available along systolic dimension 1 (columns in the array).
Notice also that the different rows of the input are mapped along systolic dimension 2 (rows in the array).
We have explicitly shown the values of the input along the systolic dimension 1 due to the modified dataflow pattern.
Unlike for depthwise convolution, the computation of \fuseconv~spans both systolic array dimensions, thereby achieving high utilization.
The span along dimension 1 increases as columns in the input increase, while the span along dimension 2 increases as rows in the input increase.
If input size is smaller than the systolic dimension $S$, then we can simultaneously map 1D convolutions across multiple channels as shown in Fig.~\ref{fig:1Dconv} (b).
Thus, \fuseconv~is a systolic algorithm which can be mapped with the modified dataflow to fully utilize both axes of a 2D systolic array.





\vspace*{-3pt}
\section{Experimental Setup and Results}
\label{sec:experimental}
\noindent 
In this section, we detail the experimental setup and share results to evaluate \fuseconv. 

\subsection{Experimental Setup}
\subsubsection{Networks}
We study 5 baseline DNNs, 4 networks from the MobileNet family (V1, V2, V3 small, and V3 large) \cite{howard2017mobilenets,sandler2018mobilenetv2, howard2019searching}, and MnasNet-B1 \cite{tan2019mnasnet}.
These networks are designed to be efficient on inference especially for edge devices, and prominently include depthwise separable convolution.
For each of these 5 networks, along with the baseline we consider 4 variants with \fuseconv.
Full and Half variants with $D = 1$ and $D = 2$ are the first two variants which replace all depthwise separable convolution layers in the baseline network with respective \fuseconv~layers. We then re-train the network with \fuseconv~layers using the setup described below. 
We consider two other variants Full-50\% and Half-50\% by using \fuseconv~replacements for only 50\% of the depthwise separable convolution layers. For 50\% variants, we do drop-in replacement for layers in such a way that maximum latency benefits are obtained.

\subsubsection{Datasets and Training Setup}
We evaluate the networks based on accuracy on ImageNet \cite{deng2009imagenet} dataset. 
We use PyTorch \cite{paszke2019pytorch} to train the models and report accuracy. Half-precision floating point (FP16) is used as the precision for both weights and activations during training and inference. 
We use standard {\em{rmsprop}} optimizer with 0.9 momentum, an initial learning rate of 0.016, with a batch size of 128 per GPU. 
The learning rate has an exponential decay of 0.97 for every 2.4 epochs. 
We maintain exponential moving averages of all weights with a decay of 0.9999, and use a weight decay of 1e-5.
We train all models on either 8 V100 or 4 P100 GPUs for 350 epochs.

\begin{table}[!h]
\centering
\vspace{0pt}
\scalebox{0.80}{
\begin{tabular}{lcccc}
\toprule
Network & \begin{tabular}[c]{@{}l@{}}ImageNet\\accuracy\end{tabular}  & \begin{tabular}[c]{@{}l@{}}MACs\\(millions)\end{tabular} & \begin{tabular}[c]{@{}l@{}}Params\\(millions)\end{tabular} & \begin{tabular}[c]{@{}l@{}}Speedup\end{tabular} \\ 
\toprule
MobileNet-V1\cite{howard2017mobilenets}              &  70.60 & 589  & 4.23 & 1x \\
MobileNet-V1 FuSe-Full                               &  72.86 & 1122 & 7.36 & 4.1x \\
MobileNet-V1 FuSe-Half                               &  72.00 & 573  & 4.20 & 6.76x \\
MobileNet-V1 FuSe-Full-50\%                          &  72.42 &  764 & 4.35 & 2.2x \\
MobileNet-V1 FuSe-Half-50\%                          &  71.77 &  578 & 4.22 & 2.36x \\
\midrule
MobileNet-V2\cite{sandler2018mobilenetv2}            & 72.00  & 315 & 3.50  & 1x  \\
MobileNet-V2 FuSe-Full                               & 72.49  & 430 & 4.46 & 5.1x \\
MobileNet-V2 FuSe-Half                               & 70.80  & 300 & 3.46 & 7.23x \\
MobileNet-V2 FuSe-Full-50\%                          & 72.11  & 361 & 3.61 & 2.0x \\
MobileNet-V2 FuSe-Half-50\%                          & 71.98  & 305 & 3.49 & 2.1x \\
\midrule
MnasNet-B1\cite{tan2019mnasnet}                      &  73.50  & 325 & 4.38 &  1x \\
MnasNet-B1 FuSe-Full                                 &  73.16  & 440 & 5.66 & 5.06x  \\
MnasNet-B1 FuSe-Half                                 &  71.48  & 305 & 4.25 & 7.15x \\
MnasNet-B1 FuSe-Full-50\%                            &  73.52  & 361 & 4.47 &  1.88x\\
MnasNet-B1 FuSe-Half-50\%                            &  72.61  & 312 & 4.35 &  1.97x\\
\midrule
MobileNet-V3 Small \cite{howard2019searching}        &  67.40  & 66 & 2.93 &  1x \\
MobileNet-V3 Small FuSe-Full                         &  67.17  & 84 & 4.44 & 3.02x \\
MobileNet-V3 Small FuSe-Half                         &  64.55  & 61 & 2.89 & 4.16x \\
MobileNet-V3 Small FuSe-Full-50\%                    &  67.91  & 73 & 3.18 & 1.6x \\
MobileNet-V3 Small FuSe-Half-50\%                    &  66.90  & 63 & 2.92 & 1.68x \\
\midrule
MobileNet-V3 Large \cite{howard2019searching}        & 75.20   & 238 & 5.47 & 1x \\
MobileNet-V3 Large FuSe-Full                         & 74.40   & 322 &10.57 & 3.61x \\
MobileNet-V3 Large FuSe-Half                         & 73.02   & 225 & 5.40 & 5.45x \\
MobileNet-V3 Large FuSe-Full-50\%                    & 74.50   & 264 & 5.57 & 1.76x \\
MobileNet-V3 Large FuSe-Half-50\%                    & 73.80   & 230 & 5.46 & 1.83x \\
\hline
\toprule
\end{tabular}
}
\caption{ImageNet performance, MACs and speedup of DNNs used to evaluate FuSeConv.}
\vspace*{-20pt}
\label{table:networks}
\end{table}

\subsubsection{Latency Estimation}
Latency is a complex function of several factors including speed of main memory, and size and speed of buffers within the systolic array.
To simplify the comparison of software choices on systolic arrays, we use the methodology formalized in SCALE-Sim \cite{samajdar2020systematic}.
Specifically, we assume that performance is limited only by operations on the systolic array: We add up the time required to load values into the array, compute in the MACs, systolically communicate partial values, and flush output beyond the array.
Since convolution operations dominate inference on DNNs and \fuseconv~is a drop-in replacement for depthwise convolution, we consider compute-bound convolutional layers (including Squeeze and Excite layers) and fully connected layers in latency estimation.
We report all the performance numbers on a $64\times64$ systolic-array. We only consider the output stationary dataflow, and add support for computing latency with proposed broadcast links.
\vspace{-5pt}

\subsection{Experimental Results}



\begin{figure*}
     \centering
     \vspace*{-20pt}
     \begin{subfigure}[b]{0.37\textwidth}
         \centering
         \includegraphics[width=\textwidth]{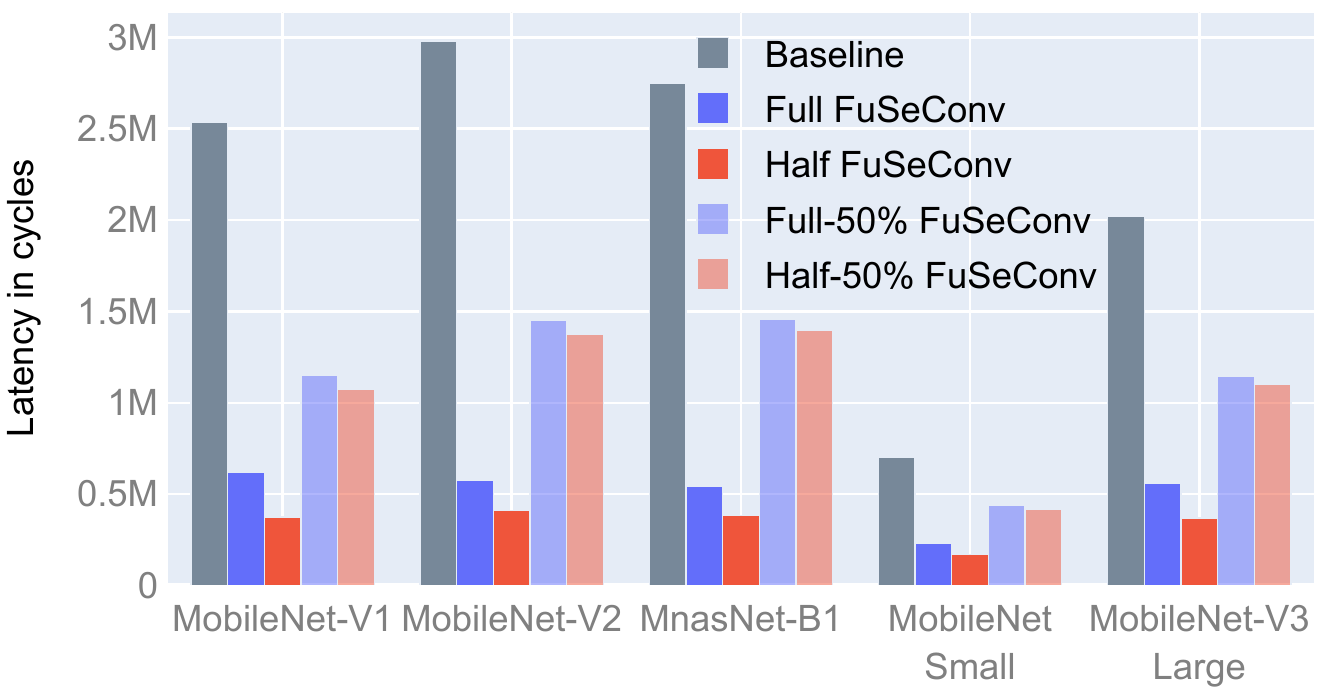}
         \caption{}
         \label{fig:res:speedup}
     \end{subfigure}
     \hfill
     \begin{subfigure}[b]{0.15\textwidth}
         \centering
         \includegraphics[width=\textwidth]{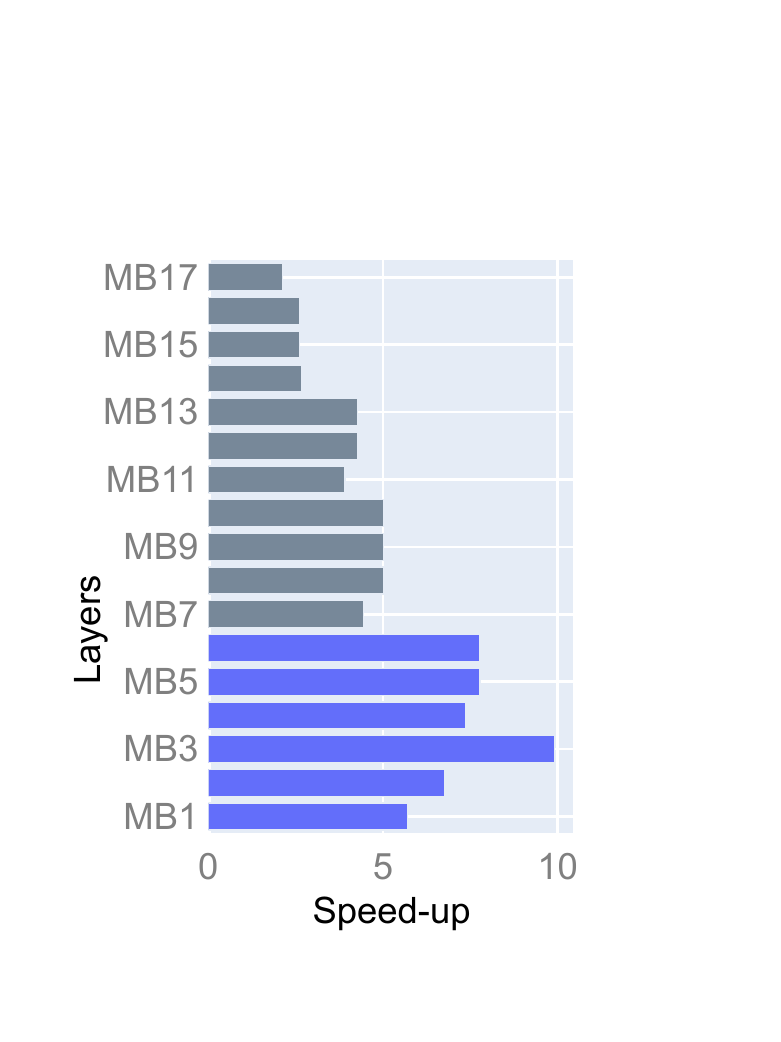}
         \caption{}
         \label{fig:res:layers}
     \end{subfigure}
     \hfill
     \begin{subfigure}[b]{0.26\textwidth}
         \centering
        \includegraphics[width=\textwidth]{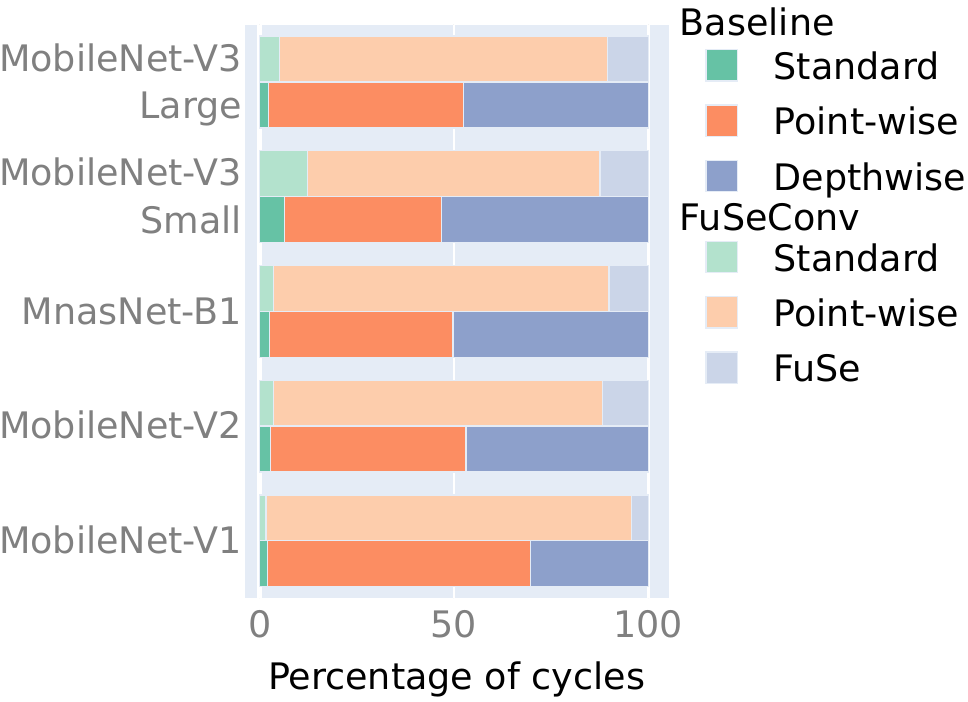}
         \caption{}
         \label{fig:res:operators}
     \end{subfigure}
     \hfill
     \begin{subfigure}[b]{0.19\textwidth}
         \centering
         \includegraphics[width=\textwidth]{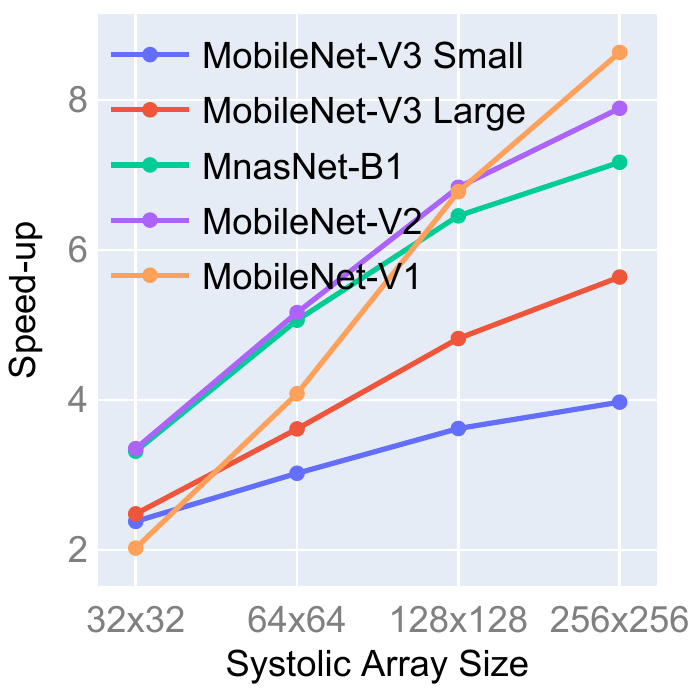}
         \caption{}
         \label{fig:res:scaling}
     \end{subfigure}
        \caption{Experimental results evaluating FuSeConv: (a) Latency estimates on $64\times64$ arrays, (b) Layer-wise speed up for MobileNetV2, (c) Latency distribution of operators for baseline and \fuseconv~networks, and (d) Ablation study}
        \label{fig:results}
    \vspace{-17pt}
\end{figure*}

%

\subsubsection{Accuracy on datasets}
We report the accuracy of different model variants in Table~\ref{table:networks}.
Note the differences between accuracy of a variant and the corresponding baseline (without \fuseconv).
For the Half variant, in 4 out of the 5 networks, there is a drop in accuracy of over 1\%. 
On the other hand, the Full variant is always within 1\% of accuracy of the baseline, with an average drop of under 0.3\%. 
This shows that the higher parameter and operation counts of the Full variant enable improved accuracy.
We thus conclude that the drop-in replacement with Full variant retains baseline accuracy.

\subsubsection{Speedup in inference time}
Table~\ref{table:networks} shows the MAC count and speed-up relative to baseline (without \fuseconv) across networks and variants. 
Additionally, Fig.~\ref{fig:results} (a) reports the exact latency of the networks.
We report significant speed-up: $4.16\times$ to $7.23\times$ with the Half variant and $3.02\times$ to $5.1\times$ with the Full variant. 
These speed-up values are relative to baseline networks designed to be efficient on the edge.
In spite of its larger MAC count, the Full variant is significantly faster than the baseline network due to the efficient mapping of \fuseconv~on to the systolic array.
The numbers for 50\% variants reveal a sensitive design trade-off between operations/latency and accuracy.


\subsubsection{Understanding the high speed-up} 
To understand the high observed speed-up, we pose two questions: Which \textit{layers} and which \textit{operators} are most responsible for the speed-up?

On the analysis of \textit{layers}, we compute layer-wise speed-up for the Full variant for MobileNetV2.
The speed-up ranges from $2.48\times$ to $9.38\times$ (see Fig.~\ref{fig:results}(b)). 
Notably, initial layers which have larger input feature maps report larger speed-up values.
This suggests that larger layers benefit more from \fuseconv~transformation due to better utilization of the systolic array.

On the analysis of \textit{operators}, we report the latency distribution of different operators before and after \fuseconv~transformation. 
As shown in Fig.~\ref{fig:results}(c), the latency of baseline networks are dominated (30-50\%) by depthwise-separable convolutions which as we saw inefficiently utilize the systolic array.
After the \fuseconv~transformation, the latency distribution drastically shifts towards point-wise convolutions while the efficient \fuseconv~operators account for a much smaller fraction (4-11\%).
This establishes that the combination of \fuseconv~and the modified dataflow significantly improves utilization.

\subsubsection{Scaling to Larger Systolic Arrays}
We study how the reported speed-up scales with increasing size of systolic arrays.
As shown in Fig.~\ref{fig:results}(d), the speed-up increases as we move to larger arrays.
This shows that the under-utilization of a systolic array becomes more stark with increasing array size.
Interestingly, under-utilization has a more severe effect on performance for larger networks.
For instance, the larger, older network MobileNetV1 shows a higher speed-up on larger arrays than the newer, smaller network MobileNetV3-small.
This suggests differentiated design for cloud and edge accelerators.



\subsubsection{Area and Power Overhead of Modified Dataflow}
To evaluate the hardware overhead of the modified dataflow, we implemented a $32\times32$ systolic-array in Bluespec System Verilog \cite{nikhil2004bluespec} and synthesized to NanGate's open cell library in 45nm. 
Two variants were implemented - with and without the weight-broadcast links.
The area and power consumption were measured using Synopsys Design Compiler.
The relative area overhead was measured to be $4.35\%$ while the power overhead was $2.25\%$.
We consider these overheads to be justifiably small considering the large speed-ups reported with \fuseconv.



\section{Conclusion}
\label{sec:conclusion}
\noindent
We analyzed why depthwise separable convolution is not efficient on systolic arrays with the formalism of regular iterative algorithms.
As a drop-in replacement, we proposed \fuseconv~which fully decomposes into 1D convolutions. 
We showed that \fuseconv~is efficiently executed on systolic arrays with a modified dataflow.
The Full variants of \fuseconv~match the accuracy of efficient networks such as MobileNet and MNasNet, but with high speed-up of about $4\times$.
Further work can explore other variants to sensitively trade-off latency and accuracy.
Also, framing \fuseconv~as the result of a manual \textit{operator} search, our work motivates automated Network Operator Search (NOS) in complement to ongoing studies on NAS. 
\section{Acknowledgements}
We thank Google Cloud Platform and Robert Bosch Centre for Data Science and Artificial Intelligence, IIT Madras for their help with compute resources for this project. We thank Gokulan for his help in modeling systolic-arrays. Finally, we thank the anonymous reviewers for their insightful comments and suggestions towards improving the work.

\bibliographystyle{IEEEtran}
\bibliography{bilbliography.bib}

\end{document}